\documentclass[
    aps,            
    prl,            
    reprint,        
    superscriptaddress, 
    floatfix,
    nofootinbib,
    preprintnumbers
]{revtex4-2}

\usepackage{graphicx}      
\usepackage{amsmath}       
\usepackage{amssymb}       
\usepackage{bm}            
\usepackage{dcolumn}       
\usepackage{xcolor}
\definecolor{DarkGreen}{rgb}{0,0.6,0}
\definecolor{DarkBlue}{rgb}{0,0,0.6}
\definecolor{gray}{RGB}{128,128,128}
\usepackage[colorlinks=true,linktocpage=true,linkcolor=DarkBlue,citecolor=DarkGreen,urlcolor=DarkGreen]{hyperref}
\usepackage[capitalize]{cleveref} 
\usepackage{orcidlink}     

\graphicspath{{figures/}}  

\definecolor{todo}{rgb}{0.80,0.00,0.00}

\newcommand{\vckm}{V_\text{CKM}}

\newcommand{\Kpnn}{K^+\to\pi^+\nu\bar{\nu}}
\newcommand{\Klnn}{K_L\to\pi^0\nu\bar{\nu}}

\newcommand{\Vcb}{|V_{cb}|}

\begin{document}

\title{KOFFEE: Kaon Observables at Future Fermilab Experimental Extensions}
\preprint{CERN-TH-2026-200, PITT-PACC-2613}
\author{Cari Cesarotti}
\email[Corresponding author: ]{carissa.cesarotti@cern.ch}
\affiliation{Theoretical Physics Department, CERN, 1211 Geneva 23, CH}

\author{Samuel Homiller}
\affiliation{Pitt PACC, Department of Physics and Astronomy, University of Pittsburgh, Pittsburgh, PA 15260, USA}


\begin{abstract}
The rare decays $K^+\to\pi^+\nu\bar\nu$ and $K_L\to\pi^0\nu\bar\nu$ are among the most precisely predicted flavor observables, enabling determinations of the CKM matrix and its CP-violating phase, as well as providing
sensitivity to new physics at scales far beyond the reach of colliders.
We propose a dedicated rare-kaon program at Fermilab, exploiting the high-intensity proton complex currently under construction for its neutrino programs and potential collider R\&D. 
In this letter, we estimate the achievable kaon flux and project the sensitivity to the aforementioned kaon golden modes, as well as the reach on new dynamics in the UV with an EFT framework.
Such a program could measure both modes with unprecedented precision, probe new physics up to $\mathcal{O}(100)$ TeV, and determine the unitarity triangle from kaon processes alone. 
All this can be achieved while sharing infrastructure with other frontier programs.
\end{abstract}

\maketitle


%
Despite the remarkable success of the LHC \cite{ATLAS:2012yve, CMS:2012qbp}, several open questions remain in the Standard Model (SM) that likely arise from new dynamics in the UV.
No compelling theory resolves these outstanding anomalies, suggesting the need for empirical inspiration at high energies.
As direct access to the energy frontier will require decades to reach in the current landscape of proposed experiments, it is essential to develop in parallel other pathways to explore high-energy physics on shorter time scales.

Another conventional approach is to indirectly test high-energy phenomena via precision measurements. 
Since new heavy particles can shift these observables from their SM predictions through off-shell or loop effects, improving precision can provide the first hint of new physics.
These deviations can manifest as inconsistencies in a global fit including many observables~\cite{Charles:2004jd, Bona:2006ah, Straub:2018kue}, which can be interpreted e.g., in an effective field theory (EFT) approach that encodes the UV dynamics in operator coefficients~\cite{Grzadkowski:2010es}.

Flavor physics---the non-trivial structure of the SM contained in the Yukawa interactions---is particularly sensitive to massive new physics~\cite{Goudzovski:2022vbt,  Aebischer:2025mwl}.
Certain rare decays and meson mixing can proceed only through flavor-changing neutral currents (FCNCs), which occur at loop-level in the SM~\cite{Glashow:1970gm, Buras:1998raa}.
These processes are therefore predicted to be small as they are suppressed by loop factors, small couplings, and cancellations from the approximate isospin symmetry between quarks.
The smallness of SM predictions has two important consequences for testing new physics (NP). 
First, the NP contribution need not be parametrically smaller than the SM process, since experimental precision beyond $\mathcal{O}(0.1\text{--}1)$ has yet to be achieved \cite{NA62:2024pjp,Chang:2026vvx, KOTO:2024zbl, LHCb:2018roe, Belle-II:2018jsg,BaBar:2014omp, BESIII:2020nme}. 
Second, if the new physics contributions do not inherit the same suppression as the SM ones, their relative smallness can be due to decoupled effects at very high energy scales, possibly orders of magnitude beyond what colliders can directly access.

In the last two decades, a suite of $B$-meson experiments have come online or concluded, providing powerful tests of the CKM paradigm.
These include LHCb~\cite{LHCb:2018roe}, Belle(-II)~\cite{Belle:2000cnh, Belle-II:2018jsg}, and BaBar~\cite{BaBar:2014omp}. 
Dedicated kaon experiments were historically crucial, especially in the discovery of CP violation~\cite{Christenson:1964fg}. 
While recent experiments have also been conducted/proposed, such as NA48/62~\cite{NA48:2002tmj, NA62:2021bji, NA62:2024pjp,Chang:2026vvx} and KOTO(-II)~\cite{KOTO:2024zbl,KOTO:2025gvq}, their constraints on the quark mixing model are no longer competitive with the $B$-system. 
Moreover, a proposed upgrade to NA62, called HIKE~\cite{Moulson:2022wrb}, was recently suspended, leaving a significant gap in the kaon program.
It is therefore timely to develop an alternative experimental strategy for kaon physics.

Kaon decays can provide excellent measurements of CP violation as well as potential new effects from heavy physics \cite{Littenberg:1989ix, Inami:1980fz}, particularly in the `golden' modes:
\begin{equation}
    K^{\pm} \rightarrow \pi^\pm \nu \bar \nu \qquad \textrm{ and } \qquad \Klnn.
\end{equation}
While the $K^+$ decay has been measured to $\mathcal{O}(20 \%)$ \cite{E949:2008btt,BNL-E949:2009dza,E949:2004uaj, Chang:2026vvx,NA62:2024pjp}, the $K_L$ mode remains experimentally undetected \cite{KOTO:2024zbl,KOTO:2020prk, KOTO:2025uqg}.
These branching ratios are extremely small---$\mathcal{O}(10^{-11})$---and theoretically well understood~\cite{Brod:2021hsj}.
The clean prediction arises from a combination of factors: the hadronic matrix element is extracted from tree-level semileptonic decay rather than calculated \cite{Isidori:2005xm}, long-distance contributions are negligible, and short-distance contributions are perturbatively calculable \cite{Inami:1980fz, Buchalla:1995vs}. 
Deviations from the SM expectation would therefore be a clear signal of new physics.

Measurements of these golden modes would provide important tests of the SM, complementary to other tree- and loop-level flavor-changing processes.
Ensuring these studies can take place is therefore a priority for the field.
In the absence of support from CERN, we must consider other siting possibilities.
While the number of sufficiently intense proton sources worldwide is limited, a promising option is a minimal upgrade of the Fermilab proton accelerator complex, as first considered in Refs.~\cite{Nagaitsev:2011kg, Holmes:2013hfa}.

Fermilab is the flagship particle physics laboratory in the United States, with significant infrastructure supporting many experiments~\cite{Muong-2:2026qnz, SeaQuest:2019hsx, Mu2e:2014fns, MicroBooNE:2015bmn}.
It is currently constructing the LBNF and DUNE~\cite{DUNE:2020lwj, DUNE:2020ypp} program, which will make definitive measurements of neutrino oscillation parameters.
Part of this construction is the needed upgrade of the proton line to support a much higher intensity beam.
There are stages of this upgrade, called the PIP-II, which has been approved~\cite{Lebedev:2015uuu, Ball:2017PIPII,2672528}, and an even more intensive upgrade called the ACE-BR~\cite{2672528}. 
Notably, the high-intensity proton source that ACE-BR would provide is also a key facilitating technology for a future muon collider (MuC), whose R\&D program requires precisely this kind of high-power proton driver to produce intense muon beams~\cite{2672528, Accettura:2023ked, Begel:2025ldu}.
Additionally, the R\&D of a MuC would enable neutrino complexes from muon storage rings for both short- (nuSTORM \cite{Delahaye:2014vvd}) and long-baseline neutrino experiments (NuMAX \cite{Delahaye:2018yfq}).
A kaon program at Fermilab could thus share infrastructure with, and help motivate, muon collider and neutrino facility development.

In this article, we propose that a minimal extension of these planned upgrades would allow Fermilab to revive the kaon program. 
The logic flows both ways: the upgrade can be expanded to support kaon physics at modest additional cost, while the kaon program itself supplies further motivation for the construction---motivation it shares with the MuC R\&D that the same high-intensity proton driver would advance.

The remainder of this article is organized as follows.
We first estimate the kaon production achievable with the upgraded Fermilab proton complex (Kaon Production), and review the relevant physics of the golden modes (Kaon Physics). 
We then present our projected reach, its implications for global CKM fits, and the sensitivity of the golden-mode predictions to new physics (Results), before discussing the broader implications of the program (Discussion).

\section{Kaon Production}\label{sec:production}

Our first goal is to determine the number of kaons realistically available for precision experiments at near-term and future upgrades to the Fermilab accelerator complex.
Following the PIP-II upgrade, the Booster will support a $8~\textrm{GeV}$ proton beam at $166~\textrm{kW}$, which leads to the $120~\textrm{GeV}$ Main Injector at $1.2~\textrm{MW}$ with a target upgrade potential of $2.4~\textrm{MW}$.
Both the Booster and Main Injector will operate at a 20 Hz repetition rate with a 53 MHz bunch train ($\sim 19$ ns bunch spacing) with $1-2$ ns bunches~\cite{Stanek:2023bae}.
A proton beam for kaon physics could potentially be delivered after either the Booster or the Main Injector acceleration. 

Contingent to the PIP-II upgrade, we consider the additional scenario for staging with NuMAX~ \cite{Delahaye:2014vvd, Delahaye:2018yfq}. 
This program utilizes protons accelerated by the linac before the Booster, achieving energies of 3--8 GeV.
We focus on two baseline beam parameters: a $3~\textrm{GeV}$, $1~\textrm{MW}$ beam for Phase-I, and a Phase-II upgrade with a $6.75~\textrm{GeV}$, $2.75~\textrm{MW}$ beam.

\begin{figure}[t!]
    \centering
    \includegraphics[width=0.95\linewidth]{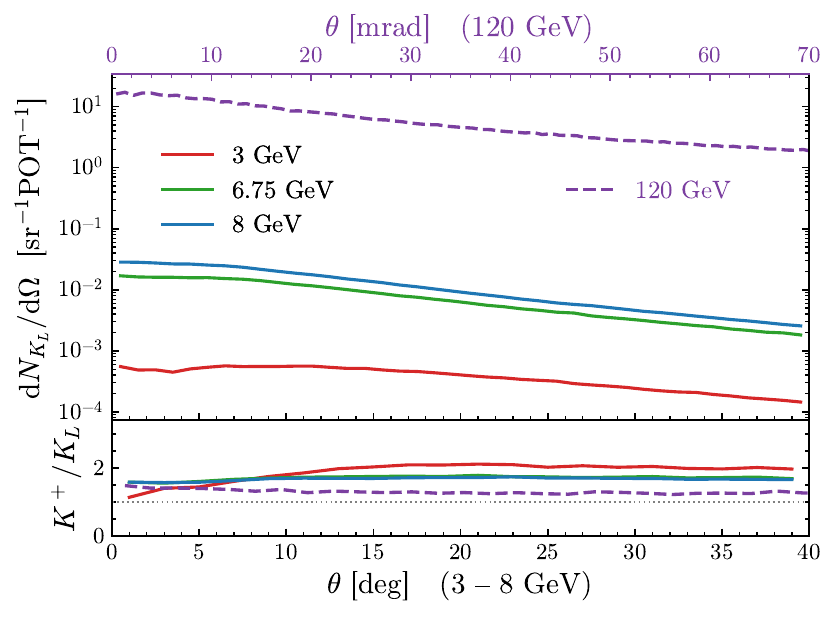}
    \vspace{-0.5em}
    \caption{The FLUKA-generated differential flux of $K_L$ in the angle $\theta$ from the beam axis.
    Note the difference in scale for high (upper) and low (lower) energy scenarios. 
    The ratio of $K^+/K_L$ per bin is also provided, and consistently 1--2$\times$ higher flux.}
    \label{fig:fluka}
\end{figure}

We estimate the kaon flux at each of these options with dedicated FLUKA~\cite{Ferrari:2005zk,Bohlen:2014buj} simulations of a proton beam on a $10~\textrm{cm}$ long, gold cylindrical target.
The differential flux of $K^+$ and $K_L$ were recorded as a function of the kaon momentum and solid angle, as shown in Fig.~\ref{fig:fluka}.
These simulations were validated by comparing to the measured charged particle and kaon flux at MIPP and NA61/SHINE~\cite{Singh:2017aro, NA61SHINE:2019aip, NA61SHINE:2022uxp}, and to the total $K_L$ yields measured by the KOTO and E391a experiments~\cite{Masuda:2015eta, Watanabe:2005gc}. 
Details of this validation, along with more simulation studies are included in the Supplementary Material.
For all the proton beam scenarios described above, the total number of kaons produced per operating year\footnote{We assume $1~\textrm{year} \approx 1.76\times 10^7~\textrm{s}$, following the live-time projections for the Main Injector program at LBNF/DUNE~\cite{DUNE:2020lwj}.} is $\sim 10^{18 - 20}$. 
%
\begin{figure*}[t!]
    \centering
    \includegraphics[width=0.8\linewidth]{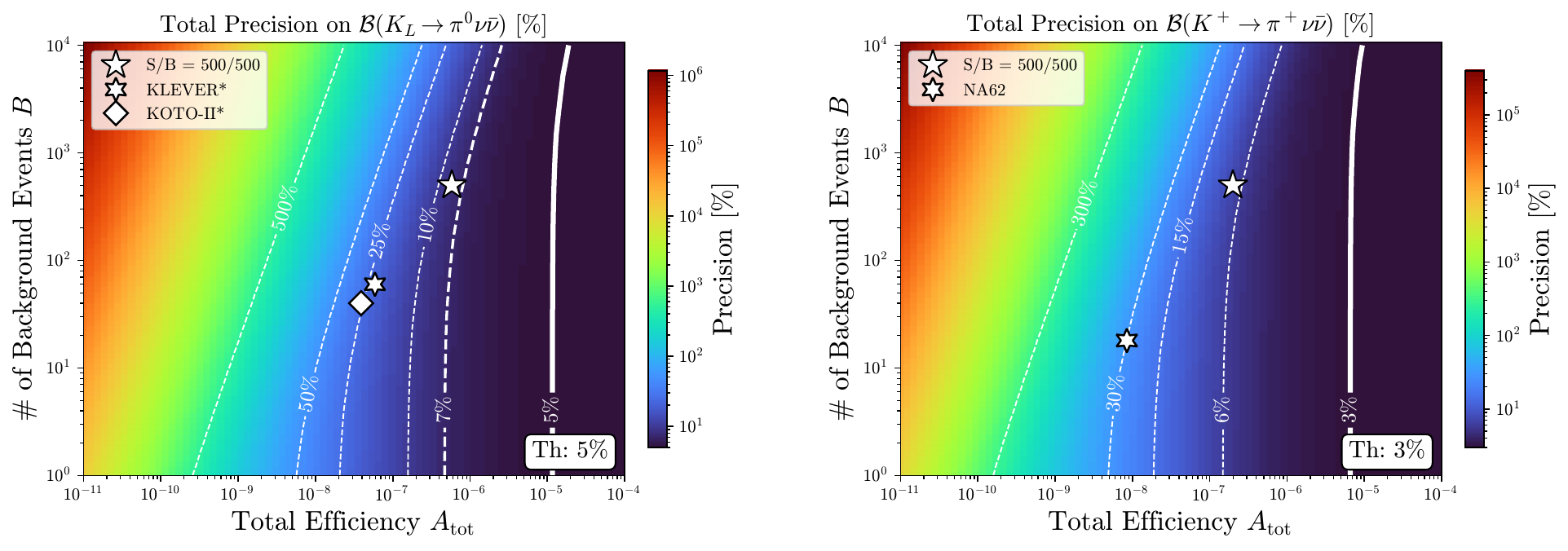}
    \caption{Contours of achievable precision for $K_L$ (left) and $K^+$ (right) golden mode decays as a function of the total efficiency $A_\text{tot}$ and number of background events $B$ for an assumed $10^{19}$ kaons per year. We show the optimistic scenario, where the theory uncertainty is halved from its current status. Beyond the thickest contour, the theory uncertainty is dominant. The plot markers show other experiments (starred labels are projections), as well as a benchmark at $S/B = 500/500$.}
    \label{fig:BgdEff}
\end{figure*}

Naturally, the sensitivity of a future experiment will depend on both the number of kaons recorded in the detector, and on the number of background events mimicking the signal of interest.
These are sensitive to a number of experimental factors that are beyond the scope of our analysis. 
Instead, taking $10^{19}$ total kaons as a benchmark, we parameterize the attainable precision on the rare kaon decays as a function of the total experimental efficiency, $A_{\textrm{tot}}$ (defined as the ratio of number of $K \to \pi\nu\nu$ decays in acceptance to the number produced, assuming the SM branching ratio) and the number of background events. 
This is illustrated in Fig.~\ref{fig:BgdEff}. 
We assume a modest improvement by a factor of two on the current theoretical uncertainties~\cite{ParticleDataGroup:2024cfk}. 
The labeled points indicate the efficiency and background numbers from existing experiments or proposals~\cite{KLEVERProject:2019aks, KOTO:2025gvq, NA62:2017rwk}.

We next estimate the geometric component of the efficiency, accounting for the number of kaon decays that occur within a fiducial volume.
We focus on $K_L$, as these cannot be steered and must be forward to be in the fiducial region, whereas a $K^+$ beam could potentially be manipulated to improve acceptance, as done for NA62~\cite{NA62:2017rwk}.
This volume is defined separately for each of the scenarios, as the kaon spectrum leads to vastly different optimal locations. 
For the 3, 6.75, and 8 GeV proton beams, the volume is defined as a $15, 25$ and $25~\textrm{m}$ long conical frustum, $3.0$, $6.4$ and $8.0~\textrm{cm}$ in opening diameter, placed $6$, $9$, and $10~\textrm{m}$ from the target, respectively, all at $5^{\circ}$ from the beam axis. 
For the $120~\textrm{GeV}$ beam from the Main Injector, the fiducial volume is $50~\textrm{m}$ long with an opening diameter of $23.5~\textrm{cm}$, sitting $120~\textrm{m}$ from the target, $8~\textrm{mrad}$ off-axis. 

The results of these estimates are summarized in Table~\ref{tab:kaon_prod}.
We see that, across all the scenarios, $\mathcal{O}(10^{13} - 10^{15})$ kaons can be produced per year in our defined fiducial volumes, as summarized in Tab.~\ref{tab:kaon_prod}. 
This leaves room for an additional $\sim 0.1 - 10^{-3}$ experimental efficiency, depending on the background values, while still permitting measurements at the level of the current theoretical precision. 
This is comparable to the efficiencies achieved by existing experiments or targeted in more detailed proposals. 
It's also worth emphasizing that this doesn't account for the beam extraction, which could take only part of the power. 
At this level, that would scale the kaon yield linearly with the beam power. 

\begin{table}
\renewcommand{\arraystretch}{1.25}
\begin{tabular}{c|ccccc}
\hline\hline
         & POT/yr             & $A_{\textrm{geo}}$    & $K_L/p^+$ & $f_{\textrm{dec}}$    & $K_L^\textrm{FV}$ / yr \\ \hline
3 GeV    & $3.7 \! \times \! 10^{22}$ & $2.3 \! \times \! 10^{-5}$ &  $4.5 \! \times \! 10^{-4}$ & $0.36$                & $1.4 \! \times \! 10^{14}$ \\
6.75 GeV & $4.4 \! \times \! 10^{22}$ & $8.4 \! \times \! 10^{-5}$ &  $7.5 \! \times \! 10^{-3}$ & $0.33$                & $9.1 \! \times \! 10^{15}$ \\
8 GeV    & $2.3 \! \times \! 10^{21}$ & $1.2 \! \times \! 10^{-4}$ &  $1.1 \! \times \! 10^{-2}$ & $0.31$                & $9.4 \! \times \! 10^{14}$ \\ \hline
120 GeV  & $1.1 \! \times \! 10^{21}$ & $1.4 \! \times \! 10^{-3}$ &  $2.9 \! \times \! 10^{-2}$ & $0.074$               & $3.3 \! \times \! 10^{15}$ \\ \hline \hline
\end{tabular}
\caption{Summary of the $K_L$ decays produced in the fiducial volumes (as defined in the text) for the different scenarios considered.
The yield of $K^+$ is be similar but slightly higher, as shown in Fig.~\ref{fig:fluka}
}
\label{tab:kaon_prod}
\end{table}

We conclude our estimate with a few comments on how the kaons might be identified. 
For the charged mode, a measurement from a high-energy beam could implement a similar strategy and instrumentation to NA62 \cite{NA62:2017rwk, NA62:2020fhy}, where a differential Cherenkov detector followed by a silicon tracker can identify the $K^+$ and the momentum.
The pion decay products can be identified with a simple drift chamber accompanied by another Cherenkov detector (also needed to mitigate the muon background).
Additionally, in the fiducial region of kaon decay, there will be calorimeter cells for photon vetoes. 
%
At lower energies, the charged kaons could be stopped and have the $\pi^+$ recoiling against missing momentum recorded in a hermitic detector, as in the E787 and E949 experiments at the AGS~\cite{E949:2007xyy}. 

The strategy for the $K_L$ decays will be slightly different. 
Again, calorimetry along the fiducial decay region is needed for photon detection, but as signal from the $\pi_0$ decay rather than background rejection. 
Existing proposals such as KOTO-II~\cite{KOTO:2025uqg}) utilize the forwardness of the kaons to increase the detection efficiency, so the measurement of the neutral golden mode would benefit from a higher energy proton beam. 

As Fermilab has hosted a multitude of experiments, many of the detector components could be recycled from other experiments, such as the CsI calorimeter crystals in Mu2e \cite{Mu2e:2014fns} or Cherenkov calorimeters from Muon g-2 \cite{Muong-2:2026qnz}. 
Additionally, the experimental equipment used by NA62 for $K^+$ detection could be transported and reused as well.

\section{Kaon Physics}
\label{sec:physics}
In this section we review the relevant kaon phenomena. 
As stated before, the processes of particular interest are the ``golden modes"---exceptionally theoretical clean modes---of kaon decay: $\Klnn,$ and $ ~\Kpnn$.
As loop-induced FCNC processes, they are dominated by short-distance top-quark dynamics, and their hadronic matrix elements are fixed by isospin from the measured $K_{e3}$ rate, permitting a determination of the branching ratios at the percent level~\cite{Buchalla:1995vs, Buras:1998raa, Buchalla:1993bv, Buras:1994ec}. 

Flavor-changing processes occur as the mass basis and the flavor basis in the quark sector are rotated by the CKM matrix $\vckm$ \cite{Kobayashi:1973fv, Cabibbo:1963yz}. 
Unitarity of the CKM matrix ($\vckm^\dagger \vckm = \mathbb{I}$) imposes a set of relations among its elements that underlie both the GIM cancellation \cite{Glashow:1970gm} and smallness of many flavor-changing processes. 
These relationships can be geometrically interpreted as the three legs of a triangle in the complex plane, known as unitarity triangles~\cite{Jarlskog:1988ii}.
The standard unitarity triangle takes the $V_{ib}^* V_{id}^{\phantom{*}}$ elements, which can be drawn in the complex plane using the Wolfenstein parameterization~\cite{Wolfenstein:1983yz,Buras:1998raa}, with  
\begin{equation}
\bar{\rho} + i \bar{\eta} = -V_{ud}^{\phantom{*}} V_{ub}^* / (V_{cd}^{\phantom{*}} V_{cb}^*) \, .
\end{equation}
Because triangles can be built from independent combinations of observables, their agreement tests the consistency of different flavor sectors of the SM \cite{Buras:2000dm}. 

Current experimental bounds have improved enormously for $\Kpnn$ due to the NA62 experiment \cite{Chang:2026vvx,NA62:2024pjp}:
\begin{equation}
    \mathcal{B}(\Kpnn) = 9.6^{+1.9}_{-1.8} \times 10^{-11}
\end{equation}
The uncertainty nonetheless remains at the $\mathcal{O}(20\%)$ level, still far from the few-percent theoretical precision.

The neutral mode remains unobserved. The current KOTO bound \cite{KOTO:2024zbl,KOTO:2020prk, KOTO:2025uqg},
\begin{equation}
    \mathcal{B}(\Klnn) < 2.2 \times 10^{-9} \quad (90\%~\mathrm{CL}),
\end{equation}
still lies roughly two orders of magnitude above the SM prediction $\mathcal{B}(\Klnn)_{\rm SM} \approx 3 \times 10^{-11}$, leaving the theoretically cleanest kaon mode as an open experimental target for KOTO$-$II \cite{KOTO:2025gvq, KOTO:2025uqg}.

The $\Klnn$ mode is purely CP-violating and has a rate $\mathcal{B}(\Klnn)\propto\bar\eta^2$, thus fixing the height of the unitarity triangle \cite{ParticleDataGroup:2024cfk}.
The $\Kpnn$ decay is CP-conserving and constrains a complementary ellipse, as $\mathcal{B}(K^+ \to \pi^+ \nu\bar\nu) \propto \left[ (\sigma\bar\eta)^2 + (\rho_c - \bar\rho)^2 \right]$, again defined in the Wolfenstein parameterization.\footnote{For a detailed treatment of the calculation and interpretation of flavor observables, see Ref.~\cite{Buras:1998raa} and modern overviews in Ref.~\cite{Goudzovski:2022vbt, Aebischer:2025mwl}}
Measurement of these decays would allow us to draw the unitarity triangle with only kaon-loop processes \cite{Dery:2025pcx}, the first such determination with data. 

A further loop-level ($\Delta S = 2$) constraint on the same plane is provided by the indirect CP-violation parameter $\varepsilon_K$, which quantifies $K^0$--$\bar K^0$ mixing and is governed by short-distance box diagrams \cite{Inami:1980fz, Buras:1990fn, Herrlich:1995hh, Herrlich:1996vf}.
Its measured value~\cite{ParticleDataGroup:2024cfk} is already far more precise than the SM prediction, whose uncertainty is dominated by the parametric input $\Vcb$. 
We therefore do not treat $\varepsilon_K$ as a target observable, but include it in the fit among the loop-induced determinations of the apex.

The unitarity triangle is at present determined almost entirely by $B$-meson observables at LHCb and Belle~II, as encoded in the global CKM fits~\cite{Charles:2004jd, Bona:2006ah}. 
A determination from kaon loops alone therefore provides an independent cross-check of the same apex: agreement tests the SM, while a discrepancy would localize new physics to a specific loop sector. 
The golden modes are moreover the $s\to d$ analogue of the $b\to s\nu\bar\nu$ transition, recently observed in $B^+\to K^+\nu\bar\nu$ at Belle~II~\cite{Belle-II:2023esi}, so the two programs combined are a powerful constraint on new physics. 

\section{Results}
\textbf{Golden Mode Unitarity Triangle}. To illustrate the implications of these projections, we present several unitarity triangles constructed from various categories of flavor physics measurements: tree-level processes, $B$-loop processes, and $K$-loop processes. 
All observables are evaluated with the tree-level CKM input scheme ($|V_{us}|$, $|V_{ub}|$, $|V_{cb}|$, $\gamma$) in \texttt{flavio}~\cite{Straub:2018kue}.
At each point in $\bar{\rho}-\bar{\eta}$, the log-likelihood is evaluated with all hadronic and short-distance nuisances profiled under Gaussian priors. 
Notably $\Vcb$, which dominates the $\varepsilon_K$ and $\Kpnn$ normalizations, is profiled throughout. A resolution of the tension between inclusive and exclusive determinations in $B$-physics would further reduce the uncertainty. 
We draw $1-\sigma$ and $2-\sigma$ confidence regions around the experimental central values, which highlights the existing tension with CKM fits~\cite{UTfit:2022hsi}.
\begin{figure}[t!]
    \centering
    \includegraphics[width=0.4\textwidth]{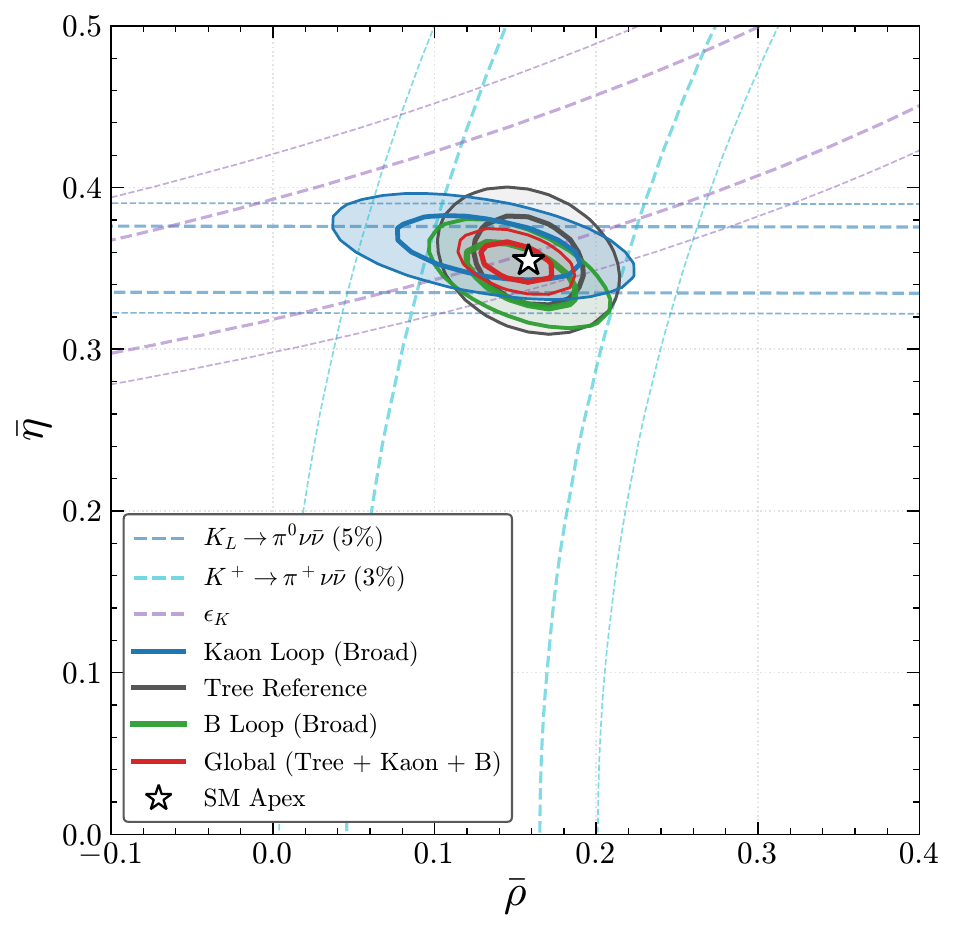}
    \vspace{-0.5em}
    \caption{The $1-$ and $2-\sigma$ confidence bands on the apex of the unitarity triangles (UT) drawn from kaon-loop measurements ($\Kpnn$, $\Klnn$, and $\epsilon_K$), tree-level ($K/B\rightarrow \pi l \nu$, $B\rightarrow D^{(*)}l \nu$, $B\rightarrow DK$), and $B$-loop ($\sin 2\beta$, $\sin 2\alpha$, $\Delta m_{d,s}$), as well as a global fit with all observables included.}
    \label{fig:UTs}
\end{figure}

The contours are shown in Fig.~\ref{fig:UTs} for both the strictly kaon-loop observables, as well as the result from combining kaon-loop, $B$-loop, and tree-level measurements. 
For the kaon golden modes, we assume that the uncertainty is dominated by a 5\% theoretical uncertainty is achieved (as is projected in the near future) to illustrate the ideal reach of the experimental program.

When we do not fix the central point to be that of the global SM fit, and instead let the experimental measurements fix the value, we see the importance of independent measurements with particular clarity. 
The kaon-loop apex is pulled to smaller values of $\bar\rho$ due to the large theory uncertainty on $\epsilon_K$, whereas the $B$-loop apex is moved only slightly by the value of $\sin 2 \beta$. 
With the first experimental determination of the kaon-loop unitarity triangle, an independent cross-check of precision flavor physics would be available. 
By increasing the number of observables in the global fit, we can more precisely evaluate the consistency of the model. 
With the distinct kaon, $B$, and tree-level processes, we improve our ability to identify the source of any potential inconsistencies, which may stem from new physics at high energies. 

\textbf{New Physics Reach.}
The experimental precision on the golden-mode branching ratio can be translated to a bound on the energy scale of new dynamics using an EFT interpretation~\cite{DAmbrosio:2002vsn, Isidori:2010kg}.
Consider a generic dimension-six operator mediating the $d_i \to d_j$ transition:
\begin{equation}
\mathcal{L} \supset\frac{c_{ij}}{\Lambda_{\rm NP}^2}\,\mathcal{O}_{ij} \equiv C^{\rm NP}_{ij}\mathcal{O}_{ij}.
\end{equation}
The SM generates the same transition at one loop, with $C^{\rm SM}_{ij} \sim (G_F\alpha/2\pi\sin^2\theta_W)\,\lambda_t^{ij} \equiv 1/\Lambda_{\rm SM}^2$, where $\lambda_t^{ij} = V_{ti}^* V_{tj}$ is CKM suppressed.
Measuring
the observable to fractional precision $p$ probes new physics down to
$C_{\rm NP} \lesssim p\,C_{\rm SM}$, i.e.
\begin{equation}
    \Lambda_{\rm NP} \;\gtrsim\; \sqrt{\frac{c_{ij}}{p}}\;\Lambda_{\rm SM}.
    \label{eq:npreach}
\end{equation}

The reach thus depends on the assumed flavor structure of the new physics.
Under minimal flavor violation (MFV)~\cite{Chivukula:1987py, DAmbrosio:2002vsn}, $c_{ij}\sim\lambda_t^{ij}$: the CKM factors cancel against those in $C_{\rm SM}$, giving a conservative reach of several TeV. 
For generic flavor violation, $c_{ij}\sim 1$, the reach is enhanced by $(1/{\lambda_t^{ij}})^{1/2}$.

We compute the corresponding bounds on the Weak Effective Theory (WET) coefficients with \texttt{flavio}~\cite{Straub:2018kue}, profiling over the SM nuisance parameters (including CKM elements, quark masses, and hadronic inputs).
Assuming precision of $p\sim0.05$--$0.1$, the reaches are summarized in Table~\ref{tab:npreach}: $K_L\to\pi^0\nu\bar\nu$ and $K^+\to\pi^+\nu\bar\nu$ probe $\sim400$ and $\sim200$~TeV respectively under generic flavor violation ($\sim8$ and $\sim4$~TeV under MFV). 
While the reach from $B$ observables is higher than that of kaons, it derives from mixing amplitudes with larger hadronic uncertainties~\cite{FlavourLatticeAveragingGroupFLAG:2024oxs, Dowdall:2019bea} compared to the cleanliness of the kaon observables. 
The kaon golden modes thus provide a theoretically clean, independent probe of new physics at energy scales beyond reach of current colliders.
 
\begin{table}[t!]
\centering
\begin{tabular}{cccc}
\hline\hline
Transition & Observable & Generic & MFV \\
\hline
$b\to d$, $\Delta F = 2$ & $\Delta M_d$ & $\sim \!2 \!\times \! 10^{3}$ TeV & $20$ TeV \\
$b\to s$, $\Delta F = 2$ & $\Delta M_s$ & $\sim \! 500$ TeV & $20$ TeV \\
$s\to d$, $\Delta F = 2$ & $\varepsilon_K$ & $\sim\!4 \! \times \! 10^{4}$~TeV & $11$~TeV  \\
$s\to d$, $\Delta F = 1$ & $\Klnn$ & $420$ TeV & $5$ TeV \\
$s\to d$, $\Delta F = 1$ & ~$\Kpnn$~ & $216$ TeV & $3.9$ TeV \\
\hline\hline
\end{tabular}
\caption{Probed new-physics scale $\Lambda$ from Eq.~\eqref{eq:npreach}, for generic ($c_{ij}\sim1$) and MFV ($c_{ij}\sim\lambda_t^{ij}$) flavor structures. 
We assume a 5\% precision on measurements.}
\label{tab:npreach}
\end{table}

\section{Discussion}
A dedicated rare-kaon program is a natural and timely addition to the Fermilab campus. 
With the PIP-II upgrade approved and under construction (and ACE-BR a potential addition), the proton source needed for a world-leading kaon facility will be shortly available. 
A modest extension of these upgrades would revive the kaon program previously proposed for CERN, while laying the groundwork for the high-power proton driver that Muon Collider R\&D demands. 
The same complex being built to drive DUNE could thus also deliver leading ultra-rare SM measurements, as well as provide the necessary infrastructure for testing the next-generation collider.

The physics case is compelling and complementary to the broader flavor program.
By improving the precision of the kaon golden-mode measurements, we could test new-physics scales well beyond what is accessible with the current collider program
Additionally, $\Klnn$ would provide a uniquely clean, purely CP-violating measurement of the CKM phase, as well as a test on the correlation to $\Kpnn$ through the Grossman--Nir bound~\cite{Grossman:1997sk}.
Realizing this program and its projected uncertainties feeds both the experimental and theoretical flavor communities; the instrumentation and analysis necessitates the expertise that might otherwise be forced to migrate from the cancellation of HIKE, and the needed improvements on the uncertainty of $\Vcb$ and  relevant lattice inputs establishes a clear target for theorists. 

The golden-mode measurements we have focused on here are not the only relevant process to study. 
An experiment designed for $K \to \pi\nu\bar{\nu}$ would inherently be sensitive to signatures such as $K \to \pi + X$~\cite{Goudzovski:2022vbt, NA62:2025upx, KOTO:2024zbl} for potential dark matter or hidden sector searches. 
Experiments could also be designed to search for other rare SM decays such as $K \to \pi^0\ell^+\ell^-$, or $K \to \mu^+\mu^-$, which provide additional complementary information on CKM unitarity~\cite{Donoghue:1987awa, Buchalla:2003sj, Dery:2021mct, Dery:2022yqc, DAmbrosio:2025mxa}. 
We encourage detailed studies of detector performance and optimization, as well as other complementary physics targets as important future work. 

A kaon program at Fermilab is a straightforward solution to many problems facing the field of particle physics. 
It could not only make the highest precision measurements on consequential SM parameters, but in doing so, indirectly reach the energy frontier, and further strengthen the need for infrastructure to run the present and future flagship particle physics experiments.

\section{Acknowledgments}
We would like to thank Patrick Meade for support and collaboration at the early stages of this project, and to him and Marzia Bordone for comments on the manuscript.
We are also grateful for useful discussions with Claudia Cornella, Roberto Franceschini, Gudrun Hiller, Gino Isidori, Sergo Jindariani, and Yotam Soreq.

\bibliographystyle{utphys}
\bibliography{refs}

\onecolumngrid
\clearpage
\begin{center}
\textbf{\large Supplemental Materials}
\end{center}
\setcounter{equation}{0}
\setcounter{figure}{0}
\setcounter{table}{0}
\setcounter{section}{0}
\setcounter{subsection}{0}
\setcounter{page}{1}

\renewcommand{\thefigure}{S\arabic{figure}}
\renewcommand{\thetable}{S\arabic{table}}
\renewcommand{\theequation}{S\arabic{equation}}
\renewcommand{\thesection}{S\arabic{section}}

\section{Details of Kaon Production}
Here we provide further details on the simulation and estimation of kaon production.
We use FLUKA~\cite{Ferrari:2005zk} version 4.5.2 to simulate the kaon yield for various proton energies and target materials.
We track the kinematic information of all outgoing particles, including other hadrons such as pions and neutrons, so taht we can estimate not only the geometric acceptance of different kaon flavors, but also validate our simulations against existing measurements and proposals.

The motivated proton energies to consider are 3 and 6.75 GeV beams from the linac, consistent with NuMAX parameters~\cite{Delahaye:2018yfq}, as well as 8~GeV and 120~GeV beams extracted from the Booster and Main Injector, respectively. 
The lower energy protons have the advantage of production rate, as low-energy protons need less acceleration and are therefore easier to attain, whereas the high-energy protons are more likely to produce kaons in the target that are boosted within geometric acceptance. 
The net production of viable kaon decays is thus non-trivial to estimate, necessitating simulation studies.

We present mainly the simulation of the $K_L$ kinematics, as the neutral-kaon golden mode is the flagship measurement to be made at this future facility, and their flux in a given fiducial volume is determined entirely by the beam, while charged Kaons can be manipulated. 
In the main text we also show the ratio per bin of $K^+$ to $K_L$ production for projections of the charged-kaon golden mode. 
Here, further present the $K^-$ flavor. 
The quark composition (namely the needed $\bar{s}$ quark) suppresses the overall flux compared to the $K^+$ production, so simulation is necessary to understand the abundance. 
The number of $K^-$ is an important consideration as the golden mode could similarly be measured by $K^- \rightarrow \pi^- \nu \bar{\nu}$. 
This measurement has not yet been made, and would be a unique probe of $CPT$ violation when compared to the $K^+$ decay.

\begin{figure*}[t!]
    \centering
    \includegraphics[width=0.45\textwidth]{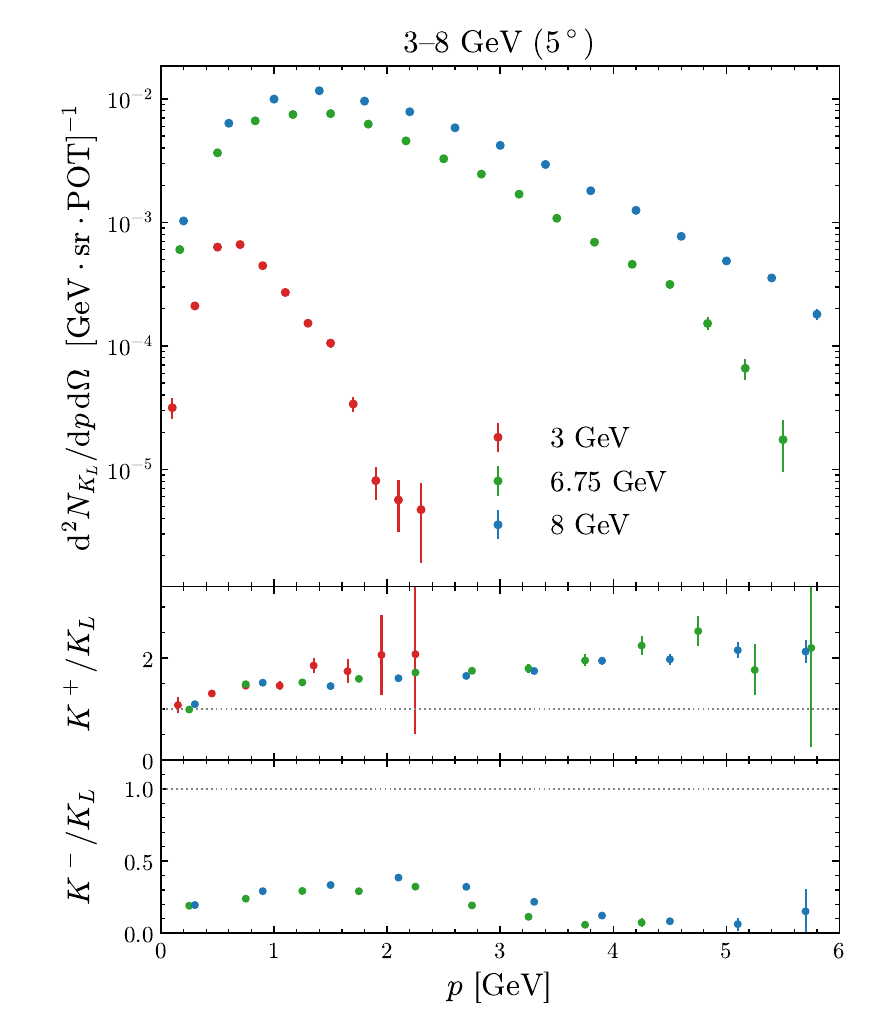}
    \qquad
    \includegraphics[width=0.45\textwidth]{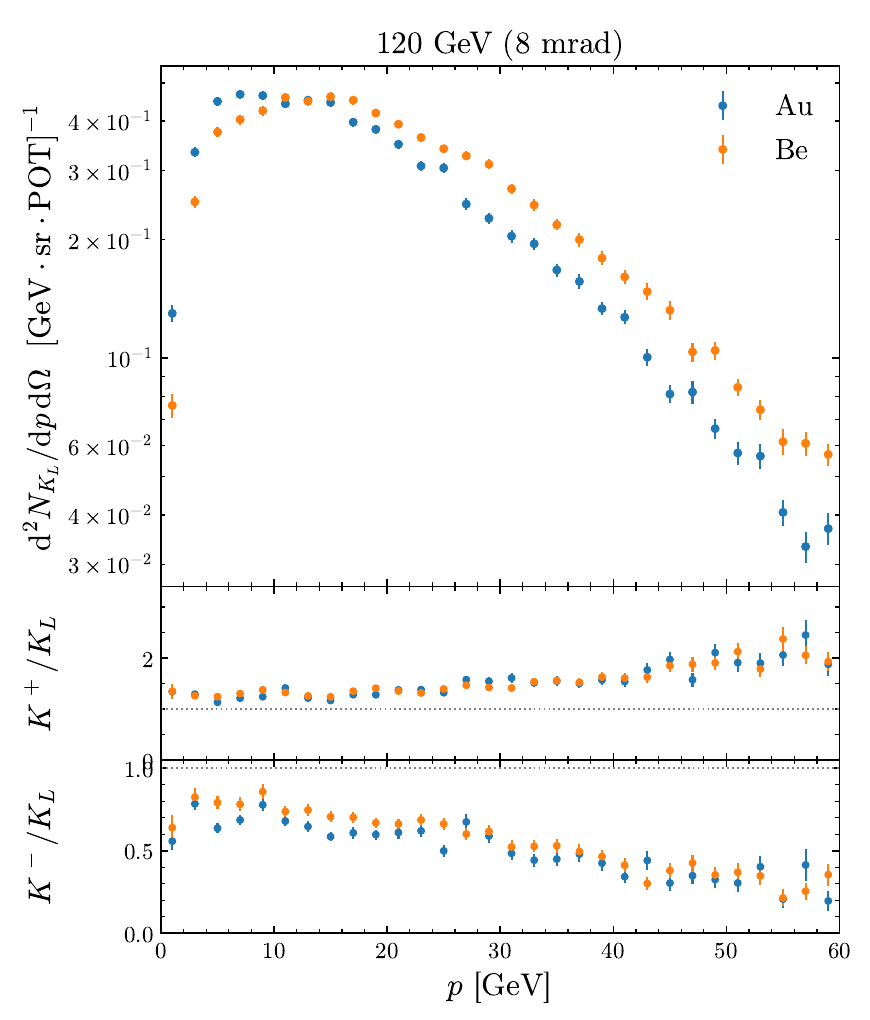}
    \caption{The momentum distribution of $K_L$ for a proton beam energy of 3, 6.75 and 8~GeV (left) and for 120 GeV, comparing beryllium and gold targets (right). The bottom panels show the ratio of other kaon flavors to $K_L$.}
    \label{fig:flukaPdist}
\end{figure*}

Additionally, we consider two different target materials that have historically been used for kaon production: beryllium and gold.
For all energies, we consider a 10 cm cylindrical target. 
Studies of optimization of target geometry are left to future work. 
We present the kaon momentum distributions from these simulations in Fig.~\ref{fig:flukaPdist}. 
Note that the target material has a negligible effect on the resultant kinematics. 
Some summary statistics of the resultant beam structure incident on the fiducial volume are presented in Table~\ref{tab:kaon_prod_supp}.

\begin{table*}[h!]
\renewcommand{\arraystretch}{1.25}
\begin{tabular}{c|cccc | cccc}
\hline\hline
         & \,$\langle p \rangle_{K_L}$\, & $p^{\textrm{peak}}_{K_L}$\,(GeV) & $\langle p \rangle_{K^+}$ (GeV) & $\langle p \rangle_{K^-}$ (GeV) 
         & $n$/POT & $n(p > \langle p \rangle_{K_L})$/POT & $n(p > 0.78\,\textrm{GeV})$ & $n(p > 2.0\,\textrm{GeV})$ \\ \hline
3 GeV    & 0.78 & 0.55 & 0.82 & 0.79 
         & \,$2.88\times 10^{-5}$\, & $1.22\times 10^{-5}$ & $1.22 \times 10^{-5}$ & $6.69 \times 10^{-6}$ \\
6.75 GeV & 1.71 & 1.42 & 1.81 & 1.59 
         & \,$9.18\times 10^{-5}$\, & $3.45 \times 10^{-5}$ & $4.48\times 10^{-5}$ & $3.14 \times 10^{-5}$ \\
8 GeV    & 1.93 & 1.30 & 2.04 & 1.81 
         & \,$1.30\times 10^{-4}$\, & $4.84\times 10^{-5}$ & $6.74\times 10^{-5}$ & $4.84\times 10^{-5}$ \\ \hline
120 GeV  & 20.1 & 7.5 & 21.9 & 17.8 
         & \,$3.92\times 10^{-4}$\, & $3.21\times 10^{-4}$ & $3.92\times 10^{-4}$ & $3.91 \times 10^{-4}$  \\ \hline \hline
\end{tabular}
\caption{
More information on the beam components, integrated over the solid angle entering the fiducial volumes defined in the text. 
}
\label{tab:kaon_prod_supp}
\end{table*}

\begin{figure}[h!]
    \centering
    \includegraphics[width=.75\linewidth]{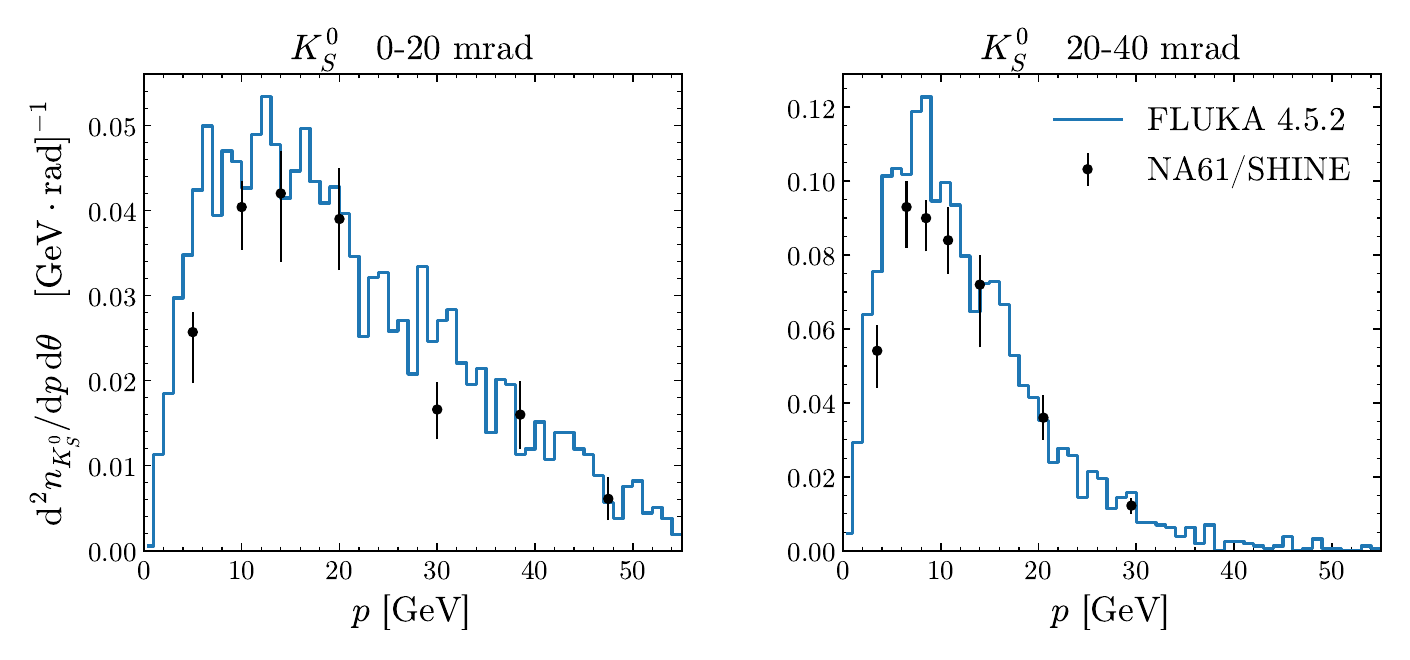}
    \vskip -0.5em
    \caption{Various validation plots comparing our FLUKA simulations to results from NA61/SHINE~\cite{NA61SHINE:2019aip, NA61SHINE:2022uxp}.
    }
    \label{fig:validate}
\end{figure}

To validate our studies, we compare the kinematics of the outgoing particle flux to results from MIPP and NA61/SHINE~\cite{Singh:2017aro, NA61SHINE:2019aip, NA61SHINE:2022uxp}. 
The results from MIPP are especially relevant as it was also a Fermilab experiment run with a 120 GeV proton beam. 
We are also able to compare the total $K_L$ yields measured by the KOTO and E391a experiments with $30$ and $12~\textrm{GeV}$ proton beams, respectively~\cite{Masuda:2015eta, Watanabe:2005gc}.  
The comparison plots are shown in in Fig.~\ref{fig:validate}
With these successful validation studies, we confidently report the number of kaons produced in acceptance within an order of magnitude. 

There are other considerations needed for a more accurate prediction of the number of usable kaon decays that are beyond the scope of this work. 
For example, in previous studies, other kaon experiments concluded that a factor of 0.3 must multiply the number of decays to account for beamline transmission through the materials of the experiment~\cite{Singh:2017aro, Watanabe:2005gc}.
Furthermore, the exact instrumentation of the experiment will change not only the acceptance but also the efficiency of keeping signal events while rejecting background. 
To understand an optimal experimental configuration for the various beam-energy possibilities as well as measurements, we encourage the involvement of the experimental community in follow-up studies.

\subsection{Neutron Beam Contamination}
We do not attempt to prescribe a full analysis strategy for the golden modes, but we remark briefly on an important beam background for the $K_L$ studies. 
The signature of a $K_L \rightarrow \pi^0 \nu \bar{\nu}$ is particularly challenging, in part because the full event cannot be reconstructed, but also because neutral particles cannot be steered, and many neutral hadrons can produce a fake signature. 
We simulate the neutron flux from our proton-on-target configurations to illustrate the importance of mitigating this background. 

As noted in Table`\ref{tab:kaon_prod_supp}, the resultant beam has more neutrons than $K_L$ as neutral particles. 
We can reduce this background by applying a cut on the angle of emission from the beam axis to reduce the neutrino halo around the forward beam. 
Kinematics alone will likely be insufficient to remove neutral-particle backgrounds--we can further reduce the probability of a neutron faking the signature by instrumenting an evacuated forward decay region (e.g. no gas for neutrons to scatter off to produce pions or other mesons).

In order to ensure the projected precision of the golden mode measurements, detailed studies of all possible backgrounds for both the charged and neutral kaons must be carried out. 
However, as the beam parameters and detector design are yet to be decided, a sophisticated study of these backgrounds is premature, and therefore excluded from this work.

\begin{figure}[t!]
    \centering
    \includegraphics[width=0.4\linewidth]{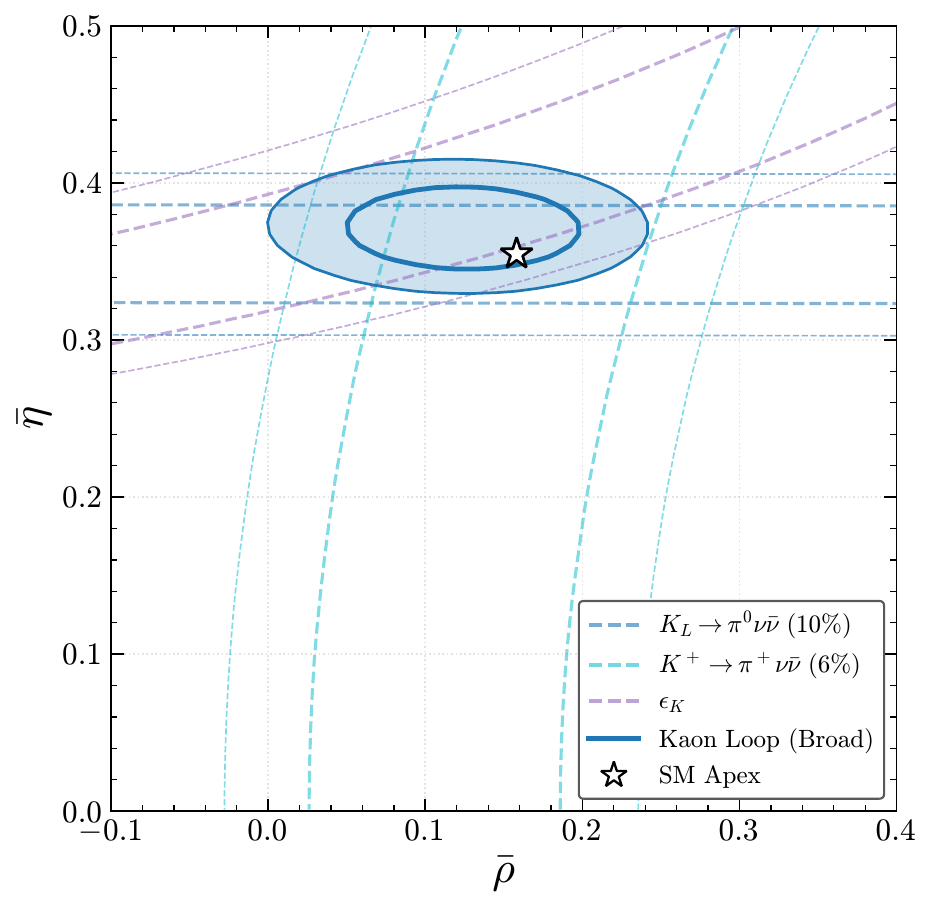}
    ~
    \includegraphics[width=0.4\linewidth]{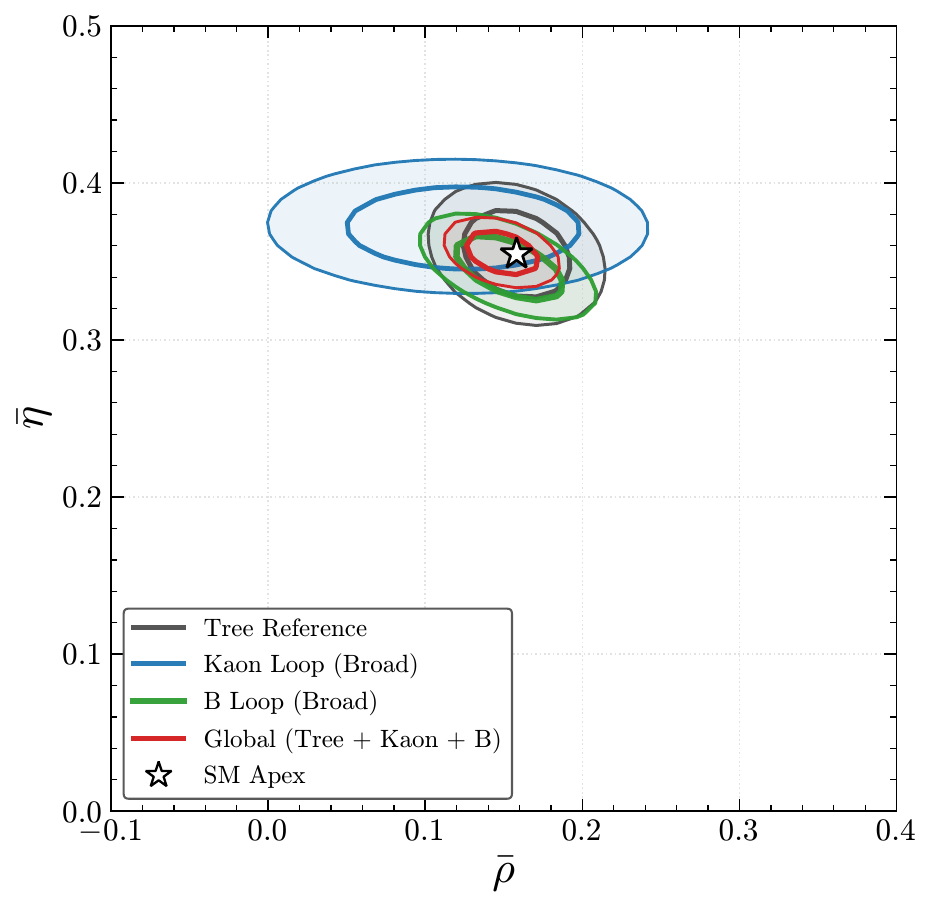}
    \caption{The $K$-loop observables (left) and global fit comparison (right) with the conservative precision for the $K$ golden-mode measurements.}
    \label{fig:conservP}
\end{figure}
\begin{figure}[t!]
    \centering
    \includegraphics[width=0.4\linewidth]{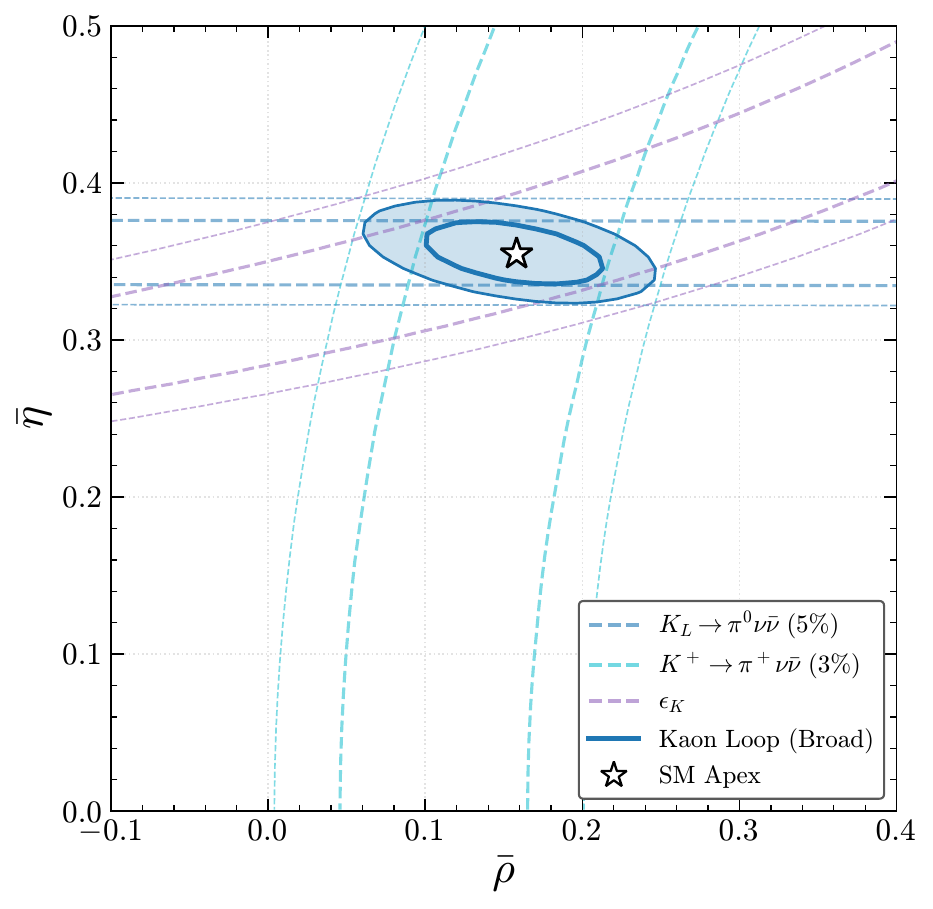}
    ~
    \includegraphics[width=0.4\linewidth]{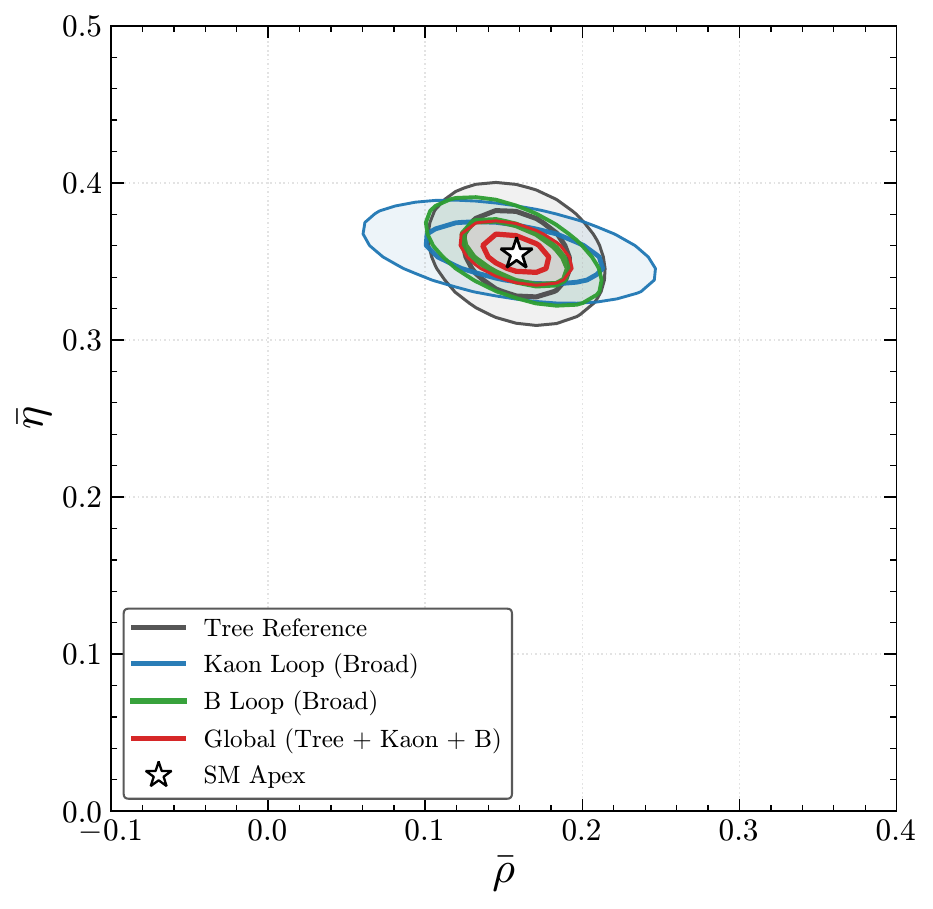}
    \caption{The $K$-loop observables (left) and global fit comparison (right) with the optimistic precision for the $K$ golden-mode measurements. The central value is defined to be the SM apex from tree-level measurements.}
    \label{fig:smApex}
\end{figure}

\section{Additional Unitarity Triangles}
In the main text we present the 1 and 2-$\sigma$ contours for both of the golden modes. 
We make two important choices for the future measurement: first, we consider an idealized uncertainty, and second, we float the apex to be determined by the parameters in the fit rather than the central value determined by the global fit over tree-level observables. 
In this section, we explore the consequences of these choices. 

We begin by revisiting the optimistic uncertainty percentage: 5\% for the $K_L$ golden mode, and 3\% for the $K_S$. 
While this is attainable, it requires significant progress in theoretical calculations in addition to the ambitious experiments we envision. 
In Fig.~\ref{fig:conservP} we present a more conservative estimate that aligns with the current state of theoretical precision: 10\% for the $K_L$ golden mode, and 6\% for the $K_S$.
For this uncertainty, it is clear that the tree-level and $B$-loop measurements provide much better precision, but the $K$-loop observable still provide a complementary constraint on CKM unitarity. 

Additionally, we consider where the apex is set. 
Thus far, we have shown plots where the central value is determined by the measured values of the observables contributing to the fit. 
The SM reference apex $(\bar\rho,\bar\eta)$ is taken from a tree-level global fit of the CKM matrix~\cite{Charles:2004jd, UTfit:2022hsi}, fixed by the new-physics-safe inputs $|V_{us}|$, $|V_{ub}|$, $|V_{cb}|$ (from semileptonic kaon and $B$ decays) and the angle $\gamma$ (from $B\to DK$).
Since we do not yet have values for the $K$-loop observables, we could also fix the central values to the SM apex in the fit. 
This largely amounts to changing the experimental value of $\varepsilon_K$ so that it agrees with the lattice predictions and tree-level inputs that enter the other unitarity triangle determinations. 
In reality, of course it is unlikely that the precise experimental value will change---an exact agreement between different determinations of the SM apex would require a combination of updated theoretical predictions and changes to other input parameters. 
Nevertheless, it is useful to consider this hypothetical scenario, in order to present the precision attainable by the full combination of different determinations, and for simplicity we continue to refer to this as the ``SM apex''. 

In Fig.~\ref{fig:smApex}, we present the results of the $K$-loop results as well as the global fit comparison centered on the SM apex.
Even when assigning the same central value, we see that the contours of the fits for the various data sets do not directly overlap. 
However, as the unitarity triangles are meant specifically to independently cross check the consistency of observables with the theoretical structure of the CKM, we argue that floating the central value is more informative. 
We emphasize again that these triangles are projections, but will change as a function of the central value and uncertainty. 

\end{document}